\documentclass{scrartcl}

\usepackage{amssymb}
\usepackage{units}
\usepackage{upgreek}
\usepackage{graphicx}
\usepackage[intoc]{nomencl}
\usepackage{pdflscape}
\usepackage{color}
\usepackage {ulem}
\usepackage{xcolor}
\usepackage[numbers,sort]{natbib}

\begin{document}

\title{Settling of spherical particles in the transitional regime}

\author{Thomas Hagemeier\thanks{
    University of Magdeburg ``Otto von Guericke", Institute of Fluid
    Dynamics and Thermodynamics, Universitaetsplatz 2, D-39106
    Magdeburg, Germany,
    Tel.: +49-391-6752427, Fax: +49-391-6742840 , Email: Thomas.Hagemeier@ovgu.de}
  \and Dominique Th\'{e}venin\thanks{
    University of Magdeburg ``Otto von Guericke", Institute of Fluid Dynamics and Thermodynamics, Universitaetsplatz 2, D-39106 Magdeburg, Germany}
  \and Thomas Richter\thanks{
    University of Magdeburg ``Otto von Guericke", Institute for
    Analysis and Numerics, Universitaetsplatz 2, D-39106 Magdeburg, Germany}
  }

\maketitle

\begin{abstract}
The settling process and wall impact of large spherical particles in a stagnant, highly viscous fluid has been observed by means of high-speed shadow imaging. The particles included in this study vary in size and material properties: steel, polytetrafluorethylen (PTFE), polyoxymethylen (POM), or rubber. The corresponding terminal Reynolds numbers range from 333 to $4012$, covering in principle the transitional  and Newton regime for drag forces. However, most particles do not reach the terminal velocity before colliding with the impact object. Therefore, the main focus of this study is set on particle settling and collision in the transitional regime. For collision studies, the Stokes number just before impact is also relevant, and lies in the range $50 < \mbox{St} < 2250$. The settling curves obtained experimentally (characterized by vertical position and vertical velocity component) are compared with numerical and analytical solutions. The latter has been derived on the basis of nominal terminal velocity and relaxation time for the Stokes and Newton regimes. The numerical model takes into account the side walls and the corresponding correction of the drag coefficient.\par 
A deviation between experimental results and analytical solution was observed in all cases where the terminal Reynolds number is larger than 300 and smaller than $1100$. It appears that, in this flow regime, the settling process of the spheres is already affected at a long distance from the impact object, leading to an early but significant deceleration. Moreover, a reduced settling velocity was observed along the whole trajectory for the PTFE particles with the lowest terminal Reynolds number. All these effects are captured in the numerical model and the corresponding results agree fairly well with the experiments. There is one exception, induced by particle rotation, which is not considered in the current model. In that case, it is not possible to correctly predict the settling process. All processed datasets are available via the Mendeley Data repository \citep{Hagemeier2020}.\par 
Two additional effects have been observed during this study. First, a bright region was detected around all PTFE spheres. It finally was found to be due to total light reflection around the sphere, but, to the best of our knowledge, this peculiarity has never been reported before. Being purely optical, this does not effect the settling behavior for PTFE. On the other hand, partial absorption of liquid at the particle surface was observed for rubber, leading to a reduced sedimentation velocity. This property, already documented for homogeneous porous particles, is found here for a particle with only an outer porous shell around a solid, impermeable core.
\end{abstract}




\section{Introduction}
\label{intro}
The situation of particles settling in a fluid has practical relevance in many industrial applications. Sometimes it is desired, e.g., to separate particles from the fluid; therefore, settling should be enhanced to accelerate the separation process. In other cases, a homogeneous distribution of particles within the fluid is important. Hence, settling must be prevented to achieve, e.g., a stable emulsion.\par
Generally speaking, particle settling is a fundamental multiphase flow process. Hence, it is not surprising to find numerous papers related to particle settling or to more complex cases like particle-wall collisions. Particle settling has been the subject of many experimental, numerical and theoretical research works (e.g., \citep{Ganji_12}, \citep{Habte_17}, \citep{Jalaal_10}, \citep{Lyotard_07}, \citep{Mordant_00}, \citep{Nouri_14}, \citep{Thompson_06}).\par
The experimental results for the particle trajectories are usually acquired with acoustic measurement methods, like in \citep{Lyotard_07} and \citep{Mordant_00}, or with optical measurements, as for instance in the seminal work of \citet{Joseph_01} and \citet{Gondret_02}. Concerning simulations, scientists attempt to develop efficient and robust modeling strategies, which would be capable of modeling accurately sedimentation as well as particle-wall collision (e.g., \citep{Habte_17}, \citep{Jenny_04}, \citep{Thompson_06}). \citet{Uhlmann_14} have provided benchmark data for single particle settling obtained from direct numerical simulation (DNS). These data cover the lower range of Reynolds numbers (Re=185 to 365) considered in the present study. \par
Recently, the interaction of elastic particles with walls has drawn attention from various researchers. Due to its complexity, this configuration is still associated to many unsolved questions. Some theoretical predictions seem to contradict reality. For instance, the contact of a
rigid particle in an incompressible fluid modeled by the linear Stokes
equations will only happen with singular forces
(e.g., \citet{Brenner_61}), but never from gravity sedimentation alone. Considering the (non-linear) Navier-Stokes equations, Feireisl~\citep{Feireisl2003} has shown that it is neither possible to establish contact, nor to release contact. These mathematical models always assume perfect incompressibility, a continuum (which, at contact and close to contact, is not meaningful); furthermore, they assume a perfect no-slip condition on
the wall and on the particle, and also perfect smoothness of all surfaces. In~\citep{GerardVaretHillairetWang2015} it has been shown that, given a free-slip non-penetration condition, contact can be established. Likewise, contact becomes possible if the surface roughness is taken into account~\citep{GerardVaretHillairet2014}, or if the fluid is considered to be compressible~\citep{Feireisl2003comp}.
All these theoretical studies analyse the case of rigid particles. If
elasticity is taken into account, the situation changes drastically. It
is assumed (but so far not proven for complex models), that for perfect
incompressibility and smoothness, no contact between the particle and
the wall will take place; still, rebouncing should be possible due to
storage and release of elastic energy \citep{Davis_86, GrandmontHillairet2016, HillairetTakahashi2009}. The rebounce
height mainly depends on the Stokes number.

The challenge for numerical studies considering particle sedimentation and
in particular particle-wall contact is found in the dynamics of the
fluid-structure interaction system
\citep{Richter2017}. This is a free boundary
value problem with moving interfaces, and it requires special techniques
for resolving the particles. The traditional approach is based on a
mapping on the moving coordinate systems onto fixed ones, the so
called Arbitrary Lagrangian Eulerian (ALE) approach, see~\citep{Donea1982,HeilHazelBoyle2008,Richter2017}. This
approach, however, fails for very large deformations and in particular
if contact is to be established; contact between a particle and the
boundary changes the topology of the domain, and no mappings of
sufficient regularity exist for transforming the Navier-Stokes
equations onto a fixed domain. Here, alternative discretizations based
on Eulerian representations of fluid and particle are possible
(e.g., \citep{Dunne2006, Frei2016, Richter2013, Uhlmann_14}). These approaches, however, are still at an early development stage and
often lack robustness and accuracy. They have been complemented by analytical and semi-analytical studies, see \citep{Ganji_12}, \citep{Jalaal_10} and \citep{Nouri_14}.

These numerous publications demonstrate the importance of theoretical considerations and analytical calculations for (i) the design of revealing experiments in general, and (ii) practical applications in particular cases.\par
\citet{Davis_86} presented a theoretical and phenomenological approach for elastic spheres moving in viscous liquids. 
Main focus was on the collision process of two spheres. Assuming that one of those spheres has an infinite radius, the solution corresponds to the collision between a sphere and a wall. The authors described the displacement of viscous fluids as a process that dissipates the kinetic energy of the approaching sphere and leads to deformation of the sphere surface. Eventually, \citet{Davis_86} derived a Stokes number criterion that allows predictions for the particle collision outcome (separation/rebound, or sticking together). In this model, the Stokes number serves as measure of the kinetic energy of the particle and is defined as:
\begin{equation} 
\mbox{St} = \frac{\rho_P~v_P~d_P}{9~\mu_F}. 
\end{equation}
It relates the properties for particle inertia (density $\rho_P$, velocity $v_P$ and diameter $d_P$) to those for the viscous forces in the fluid (viscosity $\mu_F$). Obviously, there is a direct relation with the Reynolds number $\mbox{Re}=\frac{\rho_F~v_P~d_P}{\mu_F}$, given as:  
\begin{equation} 
\mbox{St} = \frac{1}{9}\frac{\rho_P}{\rho_F} \mbox{Re}. 
\end{equation}
In case the energy level exceeds a critical threshold, particles rebound.\par
Two years later, \citet{Barnocky_88} delivered an experimental verification of this theory. During their tests, the authors investigated if metal and plastic spheres rebound or stick to a quartz surface covered by a thin film of a viscous liquid. 
The variation of relevant parameters like layer thickness, viscosity, particle diameter, and elastic properties (elastic modulus, Poisson ratio) in a wide range, provided clear evidence for the hypothesis of \citet{Davis_86}. Both studies describe particles entering a layer of viscous liquid.\par
Later, \citet{Joseph_01} continued research on particle-wall collision in a configuration with fully submerged spheres. They observed that the rebound velocity depends more on the impact Stokes number than on the elastic properties of the materials. Moreover, they mention a critical Stokes number of $\mbox{St}_{crit} = 10$ below which no rebound occurs. If the Stokes number becomes larger than 500, the restitution coefficient, which is the absolute velocity ratio after and prior to the impact $e = \frac{v}{v_0}$, approaches that for dry impact condition (particle-wall collision in air). This study is particularly relevant for the settling process, as the authors describe a significant deceleration of the particles prior to the impact. They also mention the influence of particle roughness on the repeatability of particle settling and collision experiments.\par
Almost at the same time, \citet{tenCate_02} carried out particle image velocimetry (PIV) measurements for settling particles. For low Reynolds and Stokes numbers, they obtained both, the particle trajectories as well as the flow field of the viscous fluid surrounding the particles. As the Stokes number was below the energy threshold for rebound, this study considered only the particle settling process. In addition, they carried out Lattice Boltzmann simulations of the settling process. The combination of both experimental and numerical research gave way to a more detailed understanding of the complete process. Further detailed and accurate experimental results are still required and very helpful to validate multiphase flow simulations.\par  
The studies of \citet{Gondret_99} and \citet{Gondret_02} considered in detail the process when spherical particles approach a wall and collide with it. The detailed experiments described in \citet{Gondret_02} compared various materials for the sphere and the wall, associated with a wide range of Stokes numbers. Combining their results with the data coming from \citet{Joseph_01}, the authors derived a universal correlation for the restitution coefficient as function of the collision Stokes number, and called it 'master curve'. Interestingly, this curve covers all impact events for a cascade of successive rebounds until the kinetic energy is fully dissipated. Additionally, \citet{Gondret_02} commented on wall effects coming from the target (bottom) wall and surrounding (side) walls on the restitution coefficient. They showed that the bottom wall thickness has significant effect on the restitution coefficient and its effect is proportional to the ratio of the plate thickness to the particle radius. In contrast, no significant effect was observed for the side walls when changing the distance between the settling particle and the wall. Even for a ratio of only $\frac{D}{d_P} = 2$, where $D$ represents the clear diameter of the tank, no effect was observed. \par
The research of \citet{Stocchino_05} considered the influence of non-Newtonian fluid properties on the settling and collision behavior of spherical particles. Steel spheres showed an increase in restitution coefficient for impact and rebound in a shear-thinning fluid, compared to Newtonian fluids. Additionally, \citet{Stocchino_05} provided a comprehensive description of their experimental methodology, particularly concerning image processing of noisy images, proposing finally a simple and efficient cross-correlation algorithm for the detection of particle center. \par
This was not the only study on particle settling and particle-wall collision in viscoelastic fluids. A few years later, \citet{Ardekani_09} reformulated the Stokes number, using the local strain rate, to collapse the correlations for the restitution coefficient as function of the Stokes number for Newtonian and non-Newtonian fluids. They also considered the effect of the Deborah number {$\mbox{De} = \frac{2~t_R~v_T}{d_P}$} on the collision process. According to \citet{Ardekani_09}, the Deborah number should be defined with the relaxation time $t_R$ and the collision velocity rather than with the terminal velocity of the particle, similar to the Stokes number. \par
The settling process of spherical particles not only includes the motion of the particle towards the wall and the collision mechanics, but leads also to the formation of complex vortex structures in the fluid, as shown by \citet{Thompson_07}. Here, the particle settling was well controlled by lowering the particle attached to a wire. Then, the vortex formation in the wake of the particle and the instability of the vortex structures resulting from perturbations during impact were studied. The authors compared experimental results for the vortex trajectory and maximum vorticity with numerical simulations. Their findings show axisymmetric vortex behavior at Re $< 1000$. At higher Reynolds numbers, the axial symmetry breaks down; the primary vortex becomes unstable and discrete secondary vortex rings form around the primary vortex. This complex phenomenon has been visualized in an impressive manner by \citet{Leweke_04}. \par 
A variety of more complex configurations have also attracted interest, for example the rebound of spheres from coated surfaces discussed in \citet{Davis_02}, or particles colliding with an inclined wall considered in \citet{Joseph_04}. Going even further, interesting aspects concerning contact modeling and contact time could be considered. However, this is beyond the scope of this paper and interested readers are referred to \citet{Yang_18} and \citet{Legendre_06} for a more detailed discussion.\par
Often, the analytical description of the settling process serves as basis for understanding and modeling even more complex particulate flows. The central motivation of the work presented in this paper, is to further improve our understanding of single particle settling, providing experimental data for comparison and validation of numerical work. In systematic experiments, the settling parameters (particle velocity and in-plane position as function of time) have been measured by means of particle tracking (based on shadow images) for single spheres in a water-glycerin mixture. The corresponding results will be discussed in detail in what follows. In particular, an unexpected settling behavior has been observed for the intermediate range of Reynolds numbers. Additionally, optical perturbations appeared for PTFE sphere, while a rubber sphere showed unexpectedly reduced sedimentation velocity. To the best of our knowledge, no explanation could be found in the scientific literature concerning these two observations.\par
The paper is organized as follows: Section~\ref{sec1} provides a brief overview of the settling process for spherical particles. The Stokes and Newton flow regimes and corresponding analytical solutions are derived and described in terms of relaxation time, terminal velocity, velocity-time and distance-time laws. The subsequent Section~\ref{sec2} outlines our experimental setup and measurement methodology. In Section~\ref{sec3}, results are presented and discussed. Eventually, Section~\ref{sec4} summarizes the findings of our study and describes future work.

\section{Fundamentals of particle settling}
\label{sec1}
A detailed analytical description for unsteady particle settling in different flow regimes is provided by \citet{Mann_15}. In the following, we briefly recap the laws for Stokes and Newton regimes, which are most significant for the present study. However, the full derivation is not repeated in the interest of space, and the interested reader is referred to the original research paper for more information. \par
The settling process of a single spherical particle in a stagnant fluid is governed by gravity $g$, fluid density $\rho_F$, particle material density $\rho_P$, particle diameter $d_P$, and fluid viscosity $\mu_F$. The settling or rise of a particle is the result of all forces acting on the particle in vertical direction. Therefore, further descriptions are restricted to the vertical component of all vector quantities (forces, velocities, or accelerations). \par   
Generally, the force balance considers contributions from gravitational force $F_G$, buoyancy $F_B$, drag $F_D$, inertia force of the particle $F_I$, and inertia from the surrounding fluid $F_J$. In cases where fluid density and particle density are almost equal, the contribution from particle history needs to be considered as well. If particles settle within a shear layer of a moving fluid, lift force is important too. In the present study, additional forces (history, dynamic lift, etc.) are not considered. Hence, the force balance around a spherical particle leads to \citep{Brennen_05, Crowe_06, Crowe_98}:
\begin{equation}
0 = F_G + F_B + F_D + F_I + F_J.
\end{equation} 
Rearranging the forces according to their sign, pointing in upwards or downwards direction, one obtains:
\begin{equation}
\label{eq1fun}
V_p~\rho_p~\frac{\mbox{d}v}{\mbox{d}t} + j~V_p~\rho_F~\frac{\mbox{d}v}{\mbox{d}t} = V_p~g~\left(\rho_p-\rho_F\right) + c_D~A_p~\rho_F~\frac{v^2}{2}.
\end{equation} 
The left-hand side of Eq.~\ref{eq1fun} contains the inertia forces of the particle and of the surrounding fluid, where $j$ denotes the considered amount of fluid which is additionally displaced by the particle motion. It is generally known as added mass coefficient and takes a value of $j = 0.5$ for spherical particles \citep{Brennen_05}. 
\par
On the other side appear the gravitational, buoyancy (combined in the first term) and drag forces. The variables are $V_p$ the particle volume, $\rho_p$ particle density, $\rho_F$ fluid density, $v$ particle velocity, $t$ time, $c_D$ drag coefficient and $A_p$ projection area of the particle. \par 
Various correlations are available to calculate the drag coefficient as function of the Reynolds number Re. The classical correlation of \citet{Schiller_35}:
\begin{equation}
c_D = \frac{24}{\mbox{Re}} \left(1+0.15~\mbox{Re}^{0.687}\right),
\end{equation}
valid for Re $\leq 1000$ is widely used in multiphase flow descriptions \citep{Sommerfeld_08}. The correlation proposed by \citet{Brown2003}:
\begin{equation}
c_D = \frac{24}{\mbox{Re}} \left(1+0.15~\mbox{Re}^{0.681}\right)+\frac{0.407}{1+\frac{8710}{\mbox{Re}}},
\end{equation}
is valid for a larger range, including Reynolds numbers up to Re $\leq 20,000$. Moreover, is captures the transition to the Newton regime, including the small dip of $c_D$-curve progression in the range of $1000 \leq$ Re $\leq 10,000$.

\subsection{Stokes regime}
The Stokes regime is related to a fully laminar flow around the particle and, therefore, to small Reynolds numbers $\mbox{Re}\leq 1$. The corresponding drag coefficient for a smooth sphere is 
\begin{equation}
c_D = \frac{24}{\mbox{Re}},
\end{equation}
with Re defined as
\begin{equation}
\label{eq1exp}
\mbox{Re} = \frac{v_p d_p \rho_F}{\mu_F}.
\end{equation} 
At steady-state conditions, the terminal settling velocity is
\begin{equation}
\label{eq2exp}
v_T = \frac{d^2_p~g~\left(\rho_p - \rho_F\right)}{18 \mu}.
\end{equation} 
Additionally, the relaxation time is a significant parameter in order to describe analytically particle motion. The relaxation time for a laminar flow is
\begin{equation}
\label{eq3exp}
t_R = \frac{d^2_p~\left(\rho_p+j~\rho_F\right)}{18~\mu}, 
\end{equation} 
with the same $j$ as in Eq.~(\ref{eq1fun}). Eventually, the velocity-time law for the unsteady settling of a sphere reads:
\begin{equation}
\label{eq4exp}
v(t) = v_T\left(1-\exp^{\left(-\frac{t}{t_R}\right)}\right),
\end{equation}
assuming as initial condition $v(t=0)=0~\unitfrac{m}{s}$. The corresponding distance-time law is: \par
\begin{equation}
\label{eq5exp}
s(t) = v_S\left(t-t_R\left[1-\exp^{\left(-\frac{t}{t_R}\right)}\right]\right),
\end{equation}
when the particle settling starts with $s_0=0$ at $t_0 = 0$ \citep{Mann_15}.\par

A transition is observed when the flow behavior changes from laminar to turbulent. This range is called transition regime. It corresponds to intermediate Reynolds numbers, $1 \leq \mbox{Re}\leq 1000$. There is no analytical solution for the transitional regime. 

\subsection{Newton regime}
When the Reynolds number becomes larger, $\mbox{Re}\geq 1000$, the Newton regime characterizes the turbulent flow past the particle. Here, the drag coefficient for a smooth sphere is nearly constant 
\begin{equation}
\label{eq6exp}
c_D \approx 0.44,
\end{equation}
over a large range of Re. The terminal settling velocity takes the form
\begin{equation}
\label{eq7exp}
v_T = \sqrt{\frac{3~d_p~g~\left(\rho_p - \rho_F\right)}{\rho_F}}, 
\end{equation} 
for steady-state conditions. The relaxation time is then
\begin{equation}
\label{eq8exp}
t_R = \sqrt{\frac{3~d_p~}{\rho_F~g~\left(\rho_p - \rho_F \right)}} \left(\rho_p + j~\rho_F \right). 
\end{equation} 
The final velocity-time law for the Newton regime reads:
\begin{equation}
\label{eq9exp}
v(t) = v_T~\mbox{tanh}~\left(\frac{t}{t_R}\right),
\end{equation}
assuming $v(t=0)=0~\unitfrac{m}{s}$. The associated distance-time law is:
\begin{equation}
\label{eq10exp}
s(t) = v_S\left(t_R~\ln\frac{\exp^{\left(\frac{2}{t_R}t\right)}+1}{2}-t\right),
\end{equation}
when the initial condition $s_0 = 0$ at $t_0 = 0$ is fulfilled \citep{Mann_15}.\par

The knowledge of Eqs.~(\ref{eq4exp}) and (\ref{eq5exp}) for the Stokes regime as well as Eqs.~(\ref{eq9exp}) and (\ref{eq10exp}) for the Newton regime allows to prescribe lower and upper boundaries for settling parameters (velocity and distance). These functions will be later used as reference values to analyze the experimental results. \par

\subsection{Wall effects}
A correction might be required for particles moving towards a wall, between lateral walls (inside a cylinder or rectangular-shaped container), or even near a free surface. Spheres moving in spatial confinement experience increased drag compared to Stokes law, since this law assumes an infinite distance between the particle and any boundary. \par
\citet{Brenner_61} and \citet{Tanner_63} provided corresponding wall-correction functions. Usually, as summarized by \citet{Stiess_2009}, effects coming from the side walls are corrected as a function of the ratio between wall distance and particle diameter. The end effects, observed when a settling sphere approaches a wall, need corrections with respect to the distance between the settling sphere and the wall it is approaching. These corrections are significant, for instance, for the falling-ball viscometer. However, all these corrections are only valid for low Reynolds numbers.\par
At this point, we briefly recall the corrections for the drag coefficients as provided by \citet{Brenner_61}. They are applicable for particles moving in Stokes regime. \citet{Brenner_61} describes the situation for a sphere moving towards a solid surface or a free surface. The author derives analytical solutions for Stokes regime and how they can be corrected in the vicinity of a wall or free surface. When a sphere approaches a solid wall, the increase of drag coefficient is given as correction factor ($bwc$ for bottom wall correction) of the form:
\begin{equation}
\label{eq1corrdrag}
f_{bwc} = 1 + \frac{9}{8}\frac{r}{h}.
\end{equation}
Here, $r$ is the radius of the particle and $h$ is the distance to the solid wall. In case a sphere moves towards or away from the a free surface, the correction factor ($fsc$ for free surface correction) takes the form:
\begin{equation}
\label{eq2corrdrag}
f_{fsc} = 1 + \frac{3}{4}\frac{r}{h},
\end{equation}
with $h$ being the distance of the sphere center from the free surface. However, both correction factors are valid only for small $r/h$ ratios. The inaccuracy of the correction increases significantly for larger values of $r/h$, because the correction factors represent only the first-order term of the full solution. \par
Additionally, \citet{Brenner_61} gives a correction for the drag coefficient which considers the influence of the side walls ($swc$ for side wall correction) on the motion of the sphere in the following way:
\begin{equation}
\label{eq3corrdrag}
f_{swc} = 1 + 2.105 \frac{r}{R_0}.
\end{equation}
The variable $R_0$ is the radial distance between the center of the sphere and the side wall. \par
The knowledge of these corrections is essential for a correct understanding and interpretation of experimental results. In particular, the limited spatial dimensions of many experimental setups make it unavoidable to consider the effects coming from surrounding walls. However, these correlations are not applicable to particles moving in Newton or transitional regime. Moreover, the correction coefficients cannot be superposed to obtain a global drag correction for a sphere falling, e.g., in a falling-ball viscometer, as \citet{Brenner_61} pointed out.\par
The conclusions of these considerations are, that (i) a wall correction is required for experiments carried out in spatial confinement and (ii) a modification is missing for the correction factors to be applicable outside the Stokes regime. These aspects make the wall effects an important part of the later discussion in Section \ref{sec3}.  


%
\section{Experimental configurations}
\label{sec2}
\subsection{Experimental arrangement}
The settling experiments are carried out in an acrylic-glass container filled with a water-glycerin mixture. The cylindrical container has a diameter of $D=110~\unit{mm}$ and is surrounded by another, rectangular container ($200 \times 200 ~\unit{mm}$ wide) filled with the same liquid mixture to match the refractive index in order to minimize optical distortions. The water-gylcerin mixture has a density of $\rho_F=1141~\unitfrac{kg}{m^3}$, a viscosity of $\mu_F=0.008~\unitfrac{kg}{m~s}$ and a surface tension of $\gamma= 0.064~\unitfrac{N}{m^2}$, at a volumetric mixing ratio of 0.5. All material properties have been measured in-house and compared to values known from the literature (e.g., \citet{Takamura2012} or \citet{Volk2018}).\par  

\begin{figure}
\centering
\includegraphics[width=13cm]{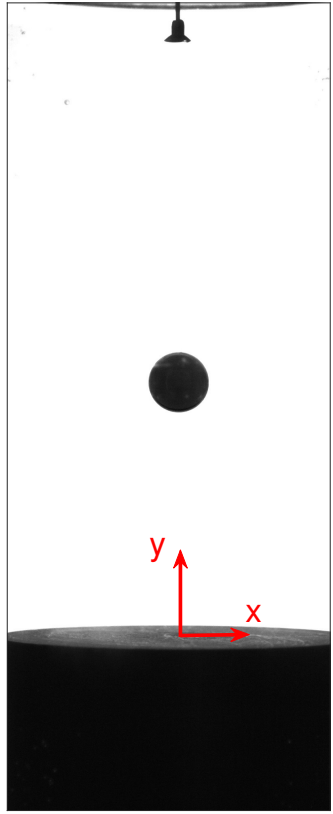}
\caption{Raw image from an experiment with a larger steel sphere showing the $(x,y)$ coordinate system. The vacuum tweezer used for holding and releasing the particle is visible at the top.}
\label{CoordSysFig1}
\end{figure}

At this point, a global coordinate system is introduced. The $x$ and $y$-coordinates are the in-plane components (see Fig.~\ref{CoordSysFig1}), while the $z$-coordinate is the out-of-plane component (not measured, and therefore not discussed further in the rest of this work), for the corresponding focus plane of the camera introduced later. The origin is defined at the upper center of a massive cylindrical slab made out of stainless steel serving as impact object, which fully covers the bottom of the acrylic-glass cylinder. The positive $y$-direction is pointing upwards. The $x$-velocity and $y$-velocity describe lateral and vertical particle motion, respectively. 
\par 

A high-speed CMOS camera captures shadow images (field of view 1268 x 2019 pixels at a scaling of 8.89 pixel/mm) of the settling process at a frame rate of $1000$ frames per second. The camera is equipped with a 60 mm objective lens and operates with a F-number of $2.8$. The camera is synchronized at the frame rate with a background illumination, consisting of several white-light LED arrays. The LED arrays are pulsed with a pulse width of 50 $\upmu s$, ensuring a homogeneous illumination throughout the exposure time of 10 $\upmu s$. \par
The spherical particles are released well below the liquid surface, and are thus from the start completely surrounded by the liquid. They are initially kept in place using a vacuum tweezer. A small cap is attached to the vacuum tweezer and holds the particles. Hence, particles are released from a small spot, without influencing the later settling process, as shown by repeating several times the experiments. The advantage of a vacuum tweezer compared to a magnetic holder is obviously that one can use any kind of test particles. It comes with the disadvantage of a delay between switching off the device and the actual particle release. However, this has no relevant consequences. The delay is taken into account by (i) additional images taken before particle motion, and (ii) a time-lag correction of the resulting trajectory and velocity curves. \par

\subsection{Material properties of test particles} 
All particles used in this study are commercially available and commonly used in ball bearings. This application ensures high production quality, in particular regarding sphericity and surface properties. The spheres are listed in Table \ref{tab3}, together with the relevant parameters such as diameter, mass, and corresponding density. 
Terminal velocities (a-priori estimates as velocity magnitude) and relaxation times are provided here as well. These values have been derived before the experiments, assuming settling in or close to the Newton regime, applying equations (\ref{eq7exp}) and (\ref{eq8exp}). They are subject to later discussions for situations where particles show strong deviations from theoretical predictions. In Table \ref{tab3}, the terminal \mbox{Re}-number and \mbox{De}-number as well as the ratio $\frac{d_p}{D}$ of particle diameter to container diameter have also been provided. 
Eventually, the Stokes number before the first impact was obtained by post-processing the measurement data and is listed in Table~\ref{tab3} as well. It is the most relevant parameter to compare the measured rebound results with existing data. The general post-processing methodology is described together with the results in Section~\ref{sec3}. 


\newpage
\begin{landscape}
\begin{table}[h!]
\caption{Particle properties}
\label{tab3}
\centering
\begin{tabular}{p{4.2cm}p{2cm}p{2cm}p{2cm}p{2cm}p{2cm}p{2cm}p{2cm}}
\hline\noalign{\smallskip}
& Steel & Steel & PTFE & PTFE & PTFE & POM & Rubber \\
\hline\noalign{\smallskip}
Diameter $d_p~\left[\unit{mm}\right]$ & 16.7 & 14.4 & 20 & 10 & 6 & 20 & 22 \\
Density $\rho_p~\left[\unitfrac{kg}{m^3}\right]$ & 7729 & 7604 & 2170 & 2158 & 2122 & 1351 & 1361 \\
Mass $m_p~\left[\unit{g}\right]$ & 18.85 & 11.89 & 9.09 & 1.13 & 0.24 & 5.66 & 7.59 \\
Terminal velocity $v_T~\left[\unitfrac{m}{s}\right]$ & 1.68 & 1.54 & 0.72 & 0.51 & 0.38 & 0.33 & 0.35 \\
Relaxation time $t_R~\left[\unit{s}\right]$ & 0.21 & 0.19 & 0.19 & 0.13 & 0.10 & 0.30 & 0.31 \\
Terminal \mbox{Re} number & 4012 & 3182 & 2078 & 730 & 333 & 939 & 1109 \\
Terminal \mbox{De} number & 42.26 & 40.64 & 13.68 & 13.26 & 12.67 & 9.00 & 8.74  \\
Diameter ratio $\frac{d_p}{D}$ & 0.15 & 0.13 & 0.18 & 0.09 & 0.055 & 0.18 & 0.2 \\
Impact \mbox{St} number & 2235 & 1698 & 387 & 114 & 53 & 70 & 118 \\
\hline
\end{tabular}
\end{table}
\end{landscape}

\newpage


%
\section{Experimental results and comparison with theory}
\label{sec3}
\subsection{Database link}
The data corresponding to these settling experiments are available via Mendeley Data \citep{Hagemeier2020}. The dataset provides for each experiment (individual repetitions) the center coordinates, the vertical and horizontal velocity components as function of time. The unit of the center coordinate values is [mm], the velocity values have the unit [m/s] and the values are given for every [ms], corresponding to a frame rate of 1000 fps. 
The ".mat"-files are a native MATLAB format. Each case includes a representative MATLAB figure. It provides a simple visualization of the particle trajectory and shows the sphere at the end of the trajectory together with an indication of the initial position. 

\subsection{Ideal particle settling}
The primary experimental result is the particle position. For each configuration, the particle position is available as function of time. The particle centroid position is the result of an image-processing algorithm that correlates a template or sample image, containing only the corresponding particle, with every other image of the series. A second-order polynomial fit is applied to the correlation peak (10 closest data points to the peak in vertical and horizontal direction), in order to obtain sub-pixel resolution and finally provide the particle centroid.\par
The particle velocity is derived from subsequent particle positions, providing in-plane (plane at focal depth) velocity components $v_x$ and $v_y$. A median filter is applied in order to reduce scattering and obtain a smooth velocity curve. \par
In addition, the reference curves for Stokes (light gray, dash-dot line) and Newton (light gray, dashed line) settling parameters are provided for each case. They are computed based on the analytical solution presented in Section~\ref{sec1}. The results below and in \ref{secA} always appear in the same order as the particles are listed from left to right in Table~\ref{tab3}. \par

\subsubsection{Steel spheres}
\begin{figure}
\centering
\includegraphics[width=13cm]{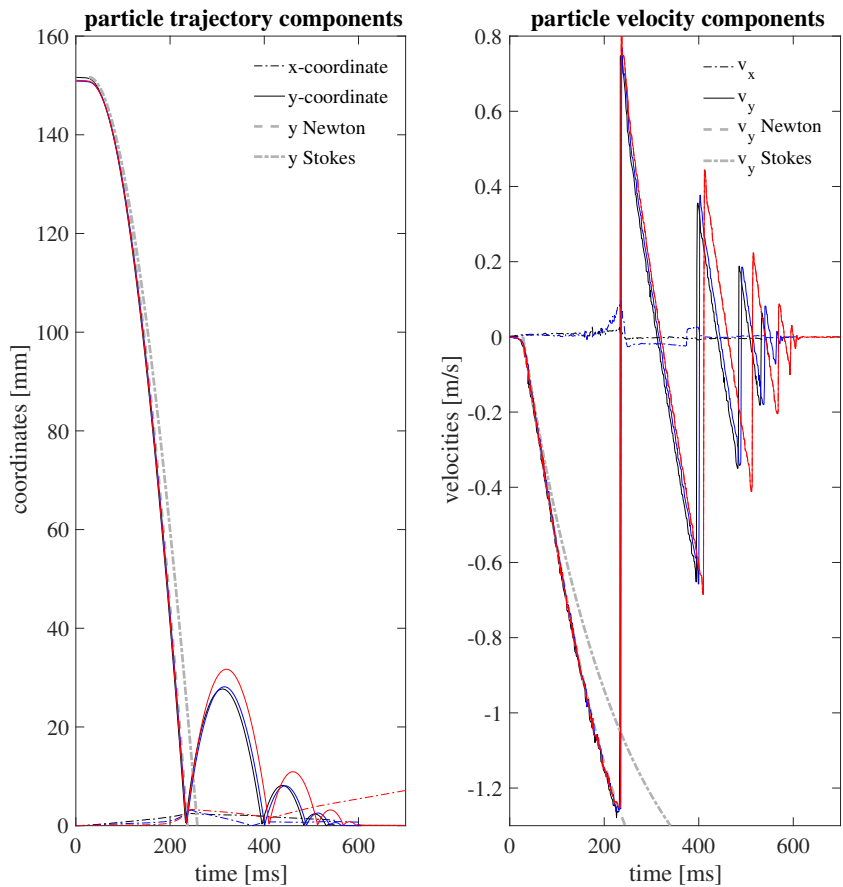}
\caption{Plots of particle trajectory (left) and velocity (right) over time for three independent settling experiments (shown with different colors -- blue, black, red) with the same steel sphere ($d_P=16.7~\unit{mm}$). Phase correction was applied to match the curves at the initial point. The lateral direction of motion can be obtained from the velocity plot, while the $x$-coordinate shows only the absolute value for better readability.}
\label{DiscFig1}
\end{figure}

Accordingly, we start the discussion with the two steel spheres (see Figs.~\ref{DiscFig1} and \ref{DiscFig2}). 
These cases correspond to the two highest Reynolds numbers. It can be seen that the settling process is identical in the initial phase for all three repetitions. The spheres accelerate throughout the settling process and collide with the impact object shortly after 200 ms for the first time. At this time, all steel spheres show almost identical velocities ($v_y\approx -1.2~\unitfrac{m}{s}$), in spite of the different diameters. Rebound occurs after the impact and in total there are five distinguishable impact/rebound events until the kinetic energy is fully dissipated.\par
The analytical solution for the Newton regime (dashed line in Fig.~\ref{DiscFig1}) describes perfectly the settling behavior. Both analytical curves, for $y$-coordinate as well as for $v_y$-velocity, collapse with the measurement data for both steel spheres and correctly predict the settling processes until the first impact. This 'ideal' settling process is completely unaffected by the experimental confinement and the measurement methodology. This confirms that the analytical description by Newton is fully appropriate for high Reynolds numbers (here, 3182 and 4012).\par

\subsection{Hindered particle settling}
In the present experiments, there is no suitable reference case for the Stokes regime, as all particles significantly exceed the limit of Stokes regime, $\mbox{Re} \leq 1$. Nevertheless, the analytical curves for the Stokes regime still deliver qualitatively correct evolutions for both $y$-coordinate and $v_y$-velocity component during the settling process. Additionally, hindered settling was observed in some experiments, for which particles occasionally show Stokes-like behavior even at Reynolds numbers $\mbox{Re} \approx 1000$ (e.g., see later Fig.~\ref{DiscFig7} and the associated discussion). Moreover, the particles usually do not reach the a-priori estimated terminal velocity. Both observations indicate an increased drag acting on particle settling in the transitional regime, as discussed in more detail in what follows. Then, the analytical solution is not appropriate any more, and a numerical solution is required. It is logical to assume that the side walls noticeably impact settling, in particular when particles are moving at low speed. In fact, all non-steel spheres do show some amount of hindered settling, as will be shown in what follows. \par

\subsubsection{Modelling of wall effect}
As a consequence, the experimental results have not only been compared with the analytical solution, but also with a numerical solution obtained for the settling velocity. For this purpose, Equation~(\ref{eq1fun}) is solved while taking into account all contributions. Additionally, the numerical model accounts for possible effects coming from free surface and bottom wall. An increased drag coefficient is obtained, according to the correction factors mentioned in equations (\ref{eq1corrdrag}) and (\ref{eq2corrdrag}). As these equations are only applicable in the Stokes regime and valid at a long distance from the wall, a prefactor $\sigma$ has been added to fit the numerical simulation using the experimental results. Eventually, the effective drag coefficient in the transitional regime is estimated in the following way:    
\begin{equation}
c_{D,eff} = c_D \cdot \left\{\sigma \cdot \max\left[\left(1+\frac{3}{4}\frac{r}{h}\right),\left(1+\frac{9}{8}\frac{r}{h}\right)\right]\right\},
\end{equation}
where the drag coefficient $c_D$ is given by the correlation of \citet{Brown2003} multiplied with the complete correction term. This correction term includes the basic correction for free surface and solid wall according to \citet{Brenner_61}. However, only the largest influence, coming from either the free surface or the solid wall is considered for each condition. This first correction is then multiplied with the empirical prefactor $\sigma$, obtained as a best fit of experimental results using the least-squares method. All prefactors found in this study are listed in Table~\ref{tab5}, together with the corrected terminal velocity $v_{T,c}$. Additionally, we provide the calculated maximum settling velocity $v_{s,max}$, which corresponds to the velocity value at the apex of the calculated velocity curve. \par   

\begin{table}[h!]
\caption{Modelling results}
\label{tab5}
\centering
\begin{tabular}{p{2.5cm}p{1cm}p{1cm}p{1cm}p{1cm}p{1cm}p{1cm}p{1.2cm}}
\hline\noalign{\smallskip}
& Steel & Steel & PTFE & PTFE & PTFE & POM & Rubber \\
\hline\noalign{\smallskip}
Diameter $d_p~\left[\unit{mm}\right]$ & 16.7 & 14.4 & 20 & 10 & 6 & 20 & 22 \\
Prefactor $\sigma$ & 0.65 & 0.7 & 0.82 & 0.89 & 0.86 & 0.89 & 0.92 \\
$v_{s,max}~\left[\unitfrac{m}{s}\right]$& 1.29 & 1.26 & 0.67 & 0.47 & 0.33 & 0.28 & 0.30 \\
$v_{T,c}~\left[\unitfrac{m}{s}\right]$& 1.64 & 1.49 & 0.68 & 0.44 & 0.30 & 0.29 & 0.31 \\
\hline
\end{tabular}
\end{table}

\subsubsection{PTFE spheres}
The PTFE-spheres show different settling behavior among each other (see Figs.~\ref{DiscFig3} to \ref{DiscFig5}), depending on the corresponding Reynolds numbers. The largest PTFE sphere ($d_P=20~\unit{mm}$) follows almost perfectly the predictions from the Newton regime regarding settling position and velocity (left and right plot of Fig.~\ref{DiscFig3}, respectively). All repetitions of this experiment show identical behavior for $y$-coordinate and $v_y$-velocity component until the first rebound event. In total, 5 impact/rebound events with significant rebound height can be identified.
The analytical description for the Newton region, appropriately predicts the settling curve. Only a minor discrepancy can be seen just before first impact, which results from the wall effects. This curve progression is better captured with our numerical model, as can be seen in Figure~\ref{DiscFig3}. However, the model slightly overestimates the maximum settling velocity. This deviation probably results from the drag correction, as only the first-order term is considered in the model. The prefactor was estimated as average for all repetitions of this configuration to $\sigma = 0.82$.  \par

The next smaller PTFE-sphere ($d_P=10~\unit{mm}$) deviates from the Newton regime prediction and shows the tendency to behave more like in the Stokes regime, see Fig.~\ref{DiscFig4}. At the difference of the previous cases, this particular sphere reaches its maximum settling velocity from Table~\ref{tab5} at around 350 ms settling time. Afterwards, the sphere decelerates until colliding with the impact object. In the present experiments, the sphere never reaches its terminal velocity, predicted prior to the measurements and listed in Table~\ref{tab3}; it stops accelerating at a velocity of $-0.47$ m/s, while being still more than 5 particle diameters away from the impact object. This is explained by the increased drag force caused by the bottom wall, since our numerical solution with a prefactor for drag correction of $\sigma = 0.89$ coincides perfectly with the measurement data. Several repetitions of the experiments show that the lateral deflection is completely insignificant for the vertical settling process (see $x$-coordinates in the left plot of Fig.~\ref{DiscFig4}). For all repetitions, the PTFE sphere decelerates exactly at the same time and distance to the bottom wall. There are three visible impact/rebound events in total, with two of them reaching a significant rebound height, before the impact energy is dissipated. \par

The smallest PTFE-sphere ($d_P=6~\unit{mm}$) shows a somewhat more classical settling behavior (see Fig.~\ref{DiscFig5}), at least concerning the shape of the velocity curve. This sphere accelerates and settles finally at a constant velocity. However, this settling velocity is noticeably lower (about 15~\%) than the terminal velocity estimated a-priori (Table~\ref{tab3}). The corrected value is given in Table~\ref{tab5}. In this case, the numerical solution slightly overestimates the influence of the drag increase caused by the presence of the wall at the end of the sedimentation trajectory; there is no significant deceleration in the experimental results, which is in contrast to the prediction of the numerical model (with a prefactor $\sigma = 0.86$). 
Still, the impact time is predicted quite correctly. Due to the low impact energy, this sphere bounces only twice, as can be seen in Fig.~\ref{DiscFig5}. Only the first rebound elevates the sphere significantly. \par

\subsubsection{POM sphere} 
The particle position obtained for the POM sphere (see Fig. \ref{DiscFig6}) shows identical progressions of the $y$-coordinate for both repetitions, during the complete settling process and even at an early stage after first rebound. Afterwards, only minor deviations in rebound height can be observed. Similar to the vertical position, the vertical velocity component matches perfectly for both tries; the two curves only start to separate when the sphere approaches the maximum rebound height for the first time.
For both cases, three rebounds can be observed, but only the first two reach significant height. \par
The settling process of the POM sphere does not fit to any of the analytical reference cases, even if the a-priori estimated terminal Reynolds number (Re=939) does suggest behavior close to Newton regime. Moreover, the settling process is not characterized by a continuous acceleration and clear convergence towards a terminal velocity value. Somewhat similar to the middle-sized PTFE sphere ($d_P = 10~\unit{mm}$), the POM sphere accelerates until reaching a maximum settling velocity at around 500 ms settling time. Afterwards, the sphere decelerates continuously, almost linearly, until colliding with the impact object. The sphere reaches a vertical velocity of about $-0.27$ m/s, far from the terminal velocity that was estimated a-priori in Table~\ref{tab3} ($-0.33$ m/s). \par
Obviously, the presence of the bottom wall increases the drag coefficient significantly. However, this effect is insufficient to fully capture the hindered settling process. In particular, the observed, linearly decreasing $v_y$ velocity component completely deviates from the model results (with best agreement found for a prefactor $\sigma = 0.89$). This suggests that the numerical model does not contain all relevant physical processes. To explain these deviations, further tests using a partly colored sphere have been conducted; these tests that the POM sphere is indeed flipped upside-down during sedimentation for all repetitions, with a total rotation of approximately $180^\circ$ around the horizontal direction. As a consequence, it is found that particle rotation (not taken into account in the analytical models, nor in the numerical model) significantly influences particle movement for POM spheres, due to a different release process from the vacuum tweezer.
After identifying the problem, these rotation tests using partly colored spheres have been systematically carried out for all materials; rotation was found insignificant in all cases, so that this problem comes only with POM spheres. Taking into account rotation in the model will be part of our future work.\par

\subsubsection{Rubber sphere}
Finally, a rubber sphere was tested. It shows (see Fig.~\ref{DiscFig7}) almost ideal Stokes-settling behavior, which was certainly not expected at first at a nominal terminal Reynolds number of $\mbox{Re}=1109$. Both, settling and velocity curves agree very well with the analytical solutions for the Stokes regime. At the end, two impact/rebound events are observed, after which the kinetic energy is dissipated.\par
After investigating this configuration in more detail, it was found that this unexpected behavior comes from the surface properties of the rubber sphere. The rubber surface is porous and absorbs some amount of liquid. In this study, all experiments have been repeated several times. However, for rubber, the sphere is only completely dry for the first settling settling experiment, with a dry connection between the sphere and the holder (vacuum tweezer). Before starting each repetition, all spheres are carefully dried with paper towels. This is sufficient for steel, PTFE, POM. However, some liquid remains in the rubber pores, as was confirmed by weighing the sphere before and after the settling experiments. This effect also causes a wet connection between the sphere and the vacuum tweezer, leading to an earlier, uncontrolled release for subsequent experiments, explaining why the very initial phase of the settling process was not captured for later experiments (see black curve in Fig.~\ref{DiscFig7}). Nevertheless, most of the settling process can still properly be captured and further analyzed without any loss of information.\par
The prediction of the numerical model including wall effects captures the experimental velocity curve reasonably well and fairly agrees with the experimental results. The best fit was obtained with a prefactor $\sigma = 0.92$. Still, small deviations can be observed, in particular for the second half of the settling process. The liquid absorption might perhaps explain these deviations and the overall settling behavior, since there is no reason to expect a Stokes regime at a terminal Reynolds number of 1109. The liquid-filled rubber probably leads to an increased frictional drag, similarly to what has been observed for homogeneous porous particles in \citep{Emadzadeh2020}. In this work, the authors report an increasing drag coefficient for porous particles in the high Re-number range ($\mbox{Re}>100$). Main difference to the work of \citep{Emadzadeh2020} is that the present rubber sphere has only a porous surface layer, with a solid metal shell below the rubber, which is very different from fully porous particles (permeation is impossible). Comparing the results and material properties of rubber and POM, there are two significant differences between both spheres, which are responsible for the completely different settling behavior in spite of almost identical characteristics (see Table \ref{tab3}). First, the surface texture is quite different, with a smooth POM sphere and a rough rubber sphere. Secondly, the moment of inertia is different. The POM sphere has an homogeneous mass distribution, while the rubber sphere is a composite of a hollow metal sphere covered by a porous, light rubber layer. Accordingly, the rubber sphere does not tend to rotate, in contrast to the POM sphere. 

%
%
\subsection{Particle-wall collision}
\begin{figure}
\centering
\includegraphics[width=12cm]{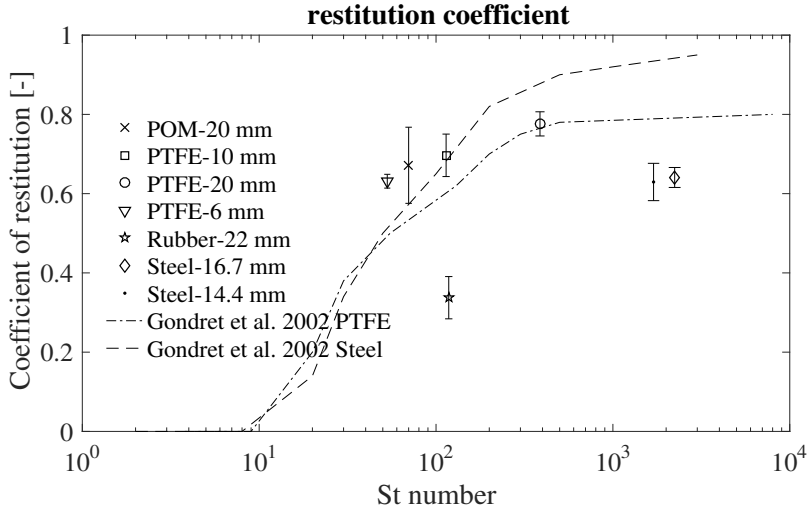}
\caption{Coefficient of restitution as function as impact Stokes number for the primary impact of various spheres on a steel target after sedimentation through a water-glycerin fluid mixture.}
\label{DiscFig8}
\end{figure}

In addition to the settling curves, the collision process has been analyzed, allowing to calculate the coefficient of restitution for the primary impact of all particles. The results are shown in Fig.~\ref{DiscFig8}. In this figure, the value of the restitution coefficient, averaged over all repetitions of the same experiment, are plotted against the corresponding impact Stokes number, listed in Table~\ref{tab3}. The representation of error bars denotes the standard deviation among the different realizations of the same experiment. In order to facilitate comparisons, the restitution curves obtained by \citet{Gondret_02} for PTFE and steel spheres are also indicated as dotted lines in this figure. \par
A relatively good agreement can be observed for the cases with PTFE spheres. Our experimental results slightly overpredict the restitution coefficient compared to \citet{Gondret_02}. Nonetheless, all trends are the same, and the differences are small. POM appears to behave similarly to PTFE in this figure. However, it must be kept in mind that particle rotation is observed for POM spheres, which is not taken into account in this computation of the restitution coefficient.\par
There is a much larger deviation between our results and those published previously concerning steel spheres. The restitution coefficients measured in our experiments are much lower than in \citet{Gondret_02}. The difference can probably be explained by the difference in impact material. While \citet{Gondret_02} used glass or glass-like walls, a massive steel block was used in our experiments. Hence, differences are not surprising.\par
The restitution coefficient of the rubber sphere is completely different. The porous rubber layer not only influences the settling process, with increased frictional drag. The elastic properties of the rubber lead to a strongly reduced restitution coefficient. The porous rubber layer, partly filled with some of the surrounding fluid, acts as damping component during the collision process and leads to enhanced impact energy dissipation.   

%
%

\subsection{Optical effect}
\begin{figure}
\centering
\includegraphics[width=8cm]{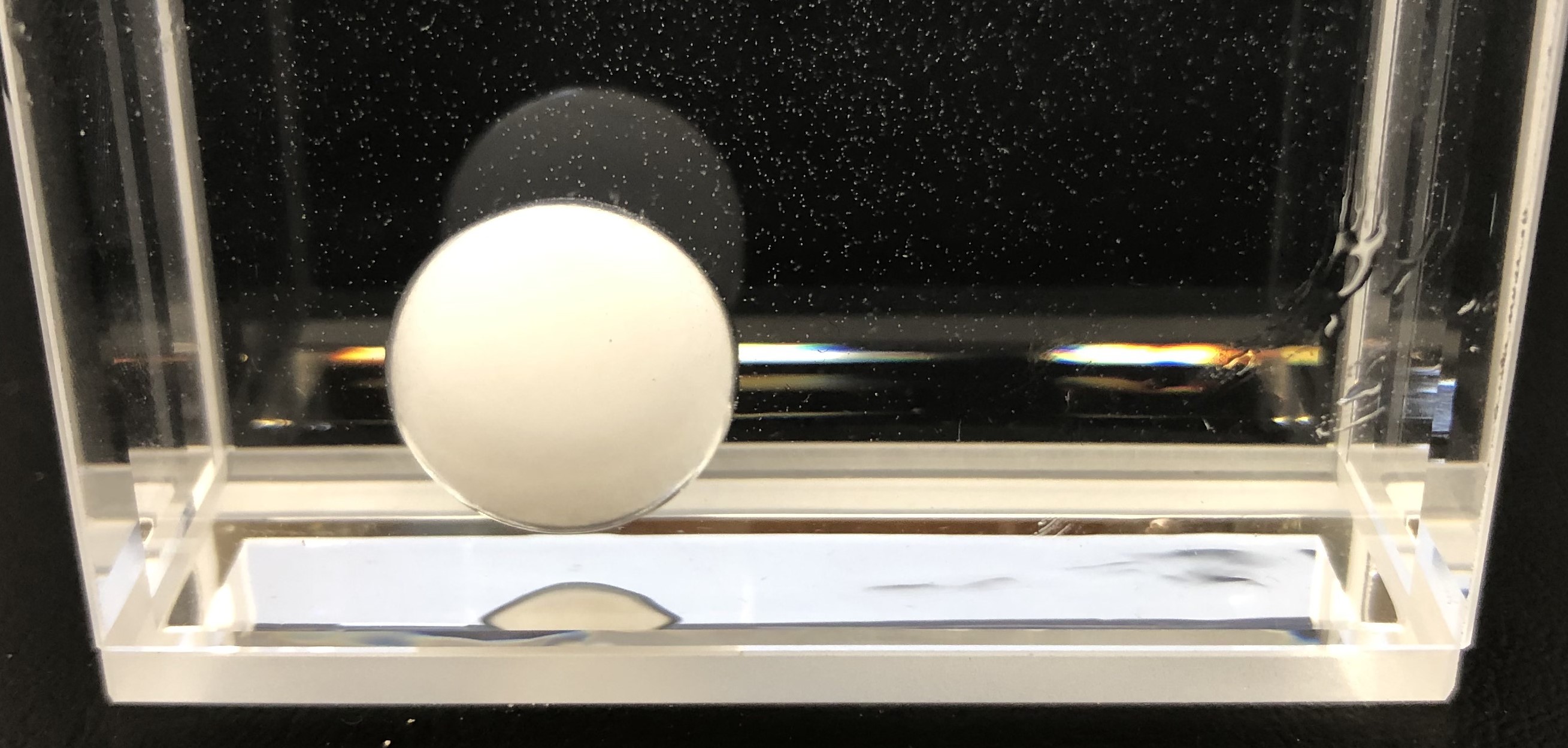}
\caption{A PTFE sphere submerged in a water-glycerin mixture at a volumetric ratio of 1:1 shows a bright halo around its surface.}
\label{DiscFig9}
\end{figure}

An unexpected optical effect was observed during all experiments with PTFE spheres. This effect is shown in Fig.~\ref{DiscFig9}, where a single PTFE sphere is imaged inside a cuvette, filled with the same water-glycerin fluid that was used for all settling experiments. Since, to the best of our knowledge, this effect has never been mentioned in the literature, a dedicated analysis has been necessary. Figure~\ref{DiscFig9} shows a bright layer around the whole sphere, best visible around the bottom half of the sphere for this particular angle of view and illumination. This effect was found to be the same in a large tank, in a small tank, and in even smaller cuvettes. It was observed identically for all PTFE spheres, independently of their size; but it was never observed for any of the other sphere materials (steel, POM, rubber).
Additional experiments have been conducted in an effort to clarify the origin of this phenomenon. First, a number of different water-glycerin mixtures with different volumetric concentrations have been prepared. High-resolution photographs of the PTFE sphere show that the thickness of the bright layer correlates directly with the glycerin concentration. A PTFE sphere in pure water shows no optical effect at all, while a PTFE sphere in pure glycerin produces the strongest effect. The intensity of the effect for a human observer is directly associated to the thickness of the layer around the sphere. The layer thickness around a PTFE sphere increases with increasing glycerin concentration, but is somewhat smaller for a larger particle diameter. At the available resolution, the total size (real particle diameter $+$ bright layer around the particle) appears to be quite constant.\par

\begin{figure}
\centering
\includegraphics[width=9cm]{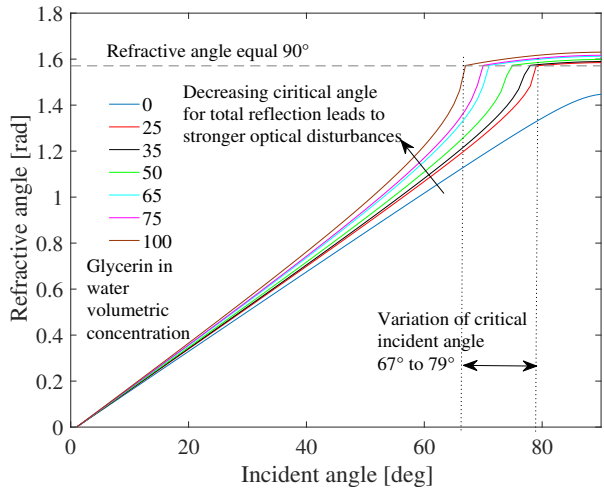}
\caption{Refractive angle as function of incident angle for several water-glycerin mixtures, for volume concentrations ranging from 0 to 100 \%.}
\label{DiscFig10}
\end{figure}

It can be concluded that the observed effect is related to a transparent layer that exists around the material sphere. More specifically, this phenomenon can be interpreted as total reflection when light passes from the medium with lower optical density (refractive index of PTFE is around $1.34$) to a medium with higher optical density (refractive index of the water-glycerin mixtures range from $1.37$ to $1.47$). Correspondingly, at the critical angle for the incident light, total reflection occurs. In Fig.~\ref{DiscFig10} the refraction angle is plotted as function of the incident angle for different water-glycerin mixtures. 
The blue curve in Fig.~\ref{DiscFig10} corresponds to pure water. This is the only curve that never leads to a total reflection, corroborating our experiments. For the water-glycerin mixtures, the refractive index increases with increasing glycerin concentration. Accordingly, the slope of the corresponding curve increases, and total reflection is observed at decreasing incident angles; the critical angle decreases from 79$^{\circ}$ at 25 \% glycerin in water to 67$^{\circ}$ for 100 \% glycerin. This effect correlates well with our experiments, where an increase of the layer thickness was observed with increasing glycerin concentration. \par
Still, the exact mechanism leading to the formation of a transparent layer, observed independently from PTFE manufacturer and particle size, could not be clarified by this additional study. A similar effect is also observed along the planar face and the corners of a PTFE plate that was used to verify the critical incident angles. Hence, the partial transparency of the PTFE surface appears to be a result of the production process. There is no reason to believe that this effect will impact settling properties, as confirmed by previous discussion. It is simply mentioned here as a help for further researchers using PTFE and glycerin mixtures.

\section{Conclusions}
\label{sec4}
During this project the settling process of spherical particles in the transitional regime has been investigated in a water-glycerin mixture. Settling curves and settling velocities have been obtained experimentally and numerically, and compared with analytical solutions for the Stokes and the Newton regimes. A good agreement with the theoretical Newton regime is observed for large steel particles at terminal Reynolds numbers above $3000$. Moreover, the collision process with a steel anvil has been investigated to obtain the corresponding restitution coefficient. \par

The study concentrates on purpose on the transitional regime, with a range of Reynolds numbers roughly between 300 and $4000$. Of particular interest are unexpected effects that appear to significantly slow down the settling process for some materials and conditions. The final settling behavior is controlled by a variety of influencing factors, some of them being well-known from previous investigations. Confirming other studies, the distance between the particle and the side walls seems to have no effect within the investigated range, since the results in a large tank are completely identical to those in a narrower one. A major influencing factor is the effect coming from the bottom wall. It significantly affects the settling curve for all particles with a-priori estimated terminal Reynolds numbers in the range of 300 to 1100. This effect leads to a changing drag coefficient. Existing drag correction approaches have been used  in combination with a pre-factor to account for the wall effects. In the absence of other influencing factors, the settling process can be predicted with a good accuracy.\par
However, this correction in the transitional settling regime fails, when other effects start to dominate, as for example significant particle rotation in the case of the POM sphere.\par

All experimental results are freely available as processed data files \cite{Hagemeier2020}, including center coordinates and vertical and horizontal velocity components. \par

An unexpected visual effect has been observed for all test particles made of PTFE, for which a bright layer appears around the whole sphere. This was finally identified as a transparent halo corresponding to the region of total light reflection. However, being purely optical, this effect has no influence on the settling process. \par

For the rubber material, liquid absorption at the surface obviously changes the material properties. This influences settling, with findings similar to those described for homogeneous porous particles. As a result, the motion of the sphere follows closely the ideal Stokes flow, and this even for a large terminal Reynolds number. Apparently, the increased frictional drag slows down the settling velocity. Understanding this effect in more detail is part of our future work. This observation could be particularly interesting for all kinds of partially-porous particles, involving a porous shell around a solid core, a feature frequently encountered for example in granulation, particle coating, and drying. Modifications of the sedimentation regime could impact industrial applications like fluidized beds. \par  

In future work, combined settling and fluid velocity measurements will be carried out. Particularly, high-speed particle image velocimetry (HS-PIV) measurements are an option to better understand the coupling of fluid and particle motion. Additionally, the combined effects from particle rotation and bottom wall on the drag coefficient appear particularly interesting. \par

The results presented so far cannot reveal the contact between the particles and the steel anvil. In particular, they do not give answers on the necessity to integrate aspects like elasticity of the solid, or compressibility of the fluid. However, these aspects could be better understood thanks to detailed mathematical models, taking such properties into account. The experimental results presented here can serve as reference data for such a model validation, since they contain both settling and multiple rebounds. 

\section{Acknowledgments}
The authors thank Dr. Schlinkert (Lab. of Mechanical Process Engineering in Magdeburg) for supporting the present research by providing technical devices, as well as Peter Kov\'{a}ts, Fabio Martins and Lena Bednarz for helpful and inspiring discussions.



\newpage
\appendix
\section{}
\label{secA}
In \ref{secA} additional results are provided regarding settling curves for different materials and different particle sizes. All data are provided without any filtering. Individual repetitions of the same experiments are shown overlaid with different colors, always in the order black, blue, red, yellow, green, and magenta. Quite often, the corresponding curves overlap, and only the last measurement curve is clearly visible. 
\begin{figure}
\centering
\includegraphics[width=13cm]{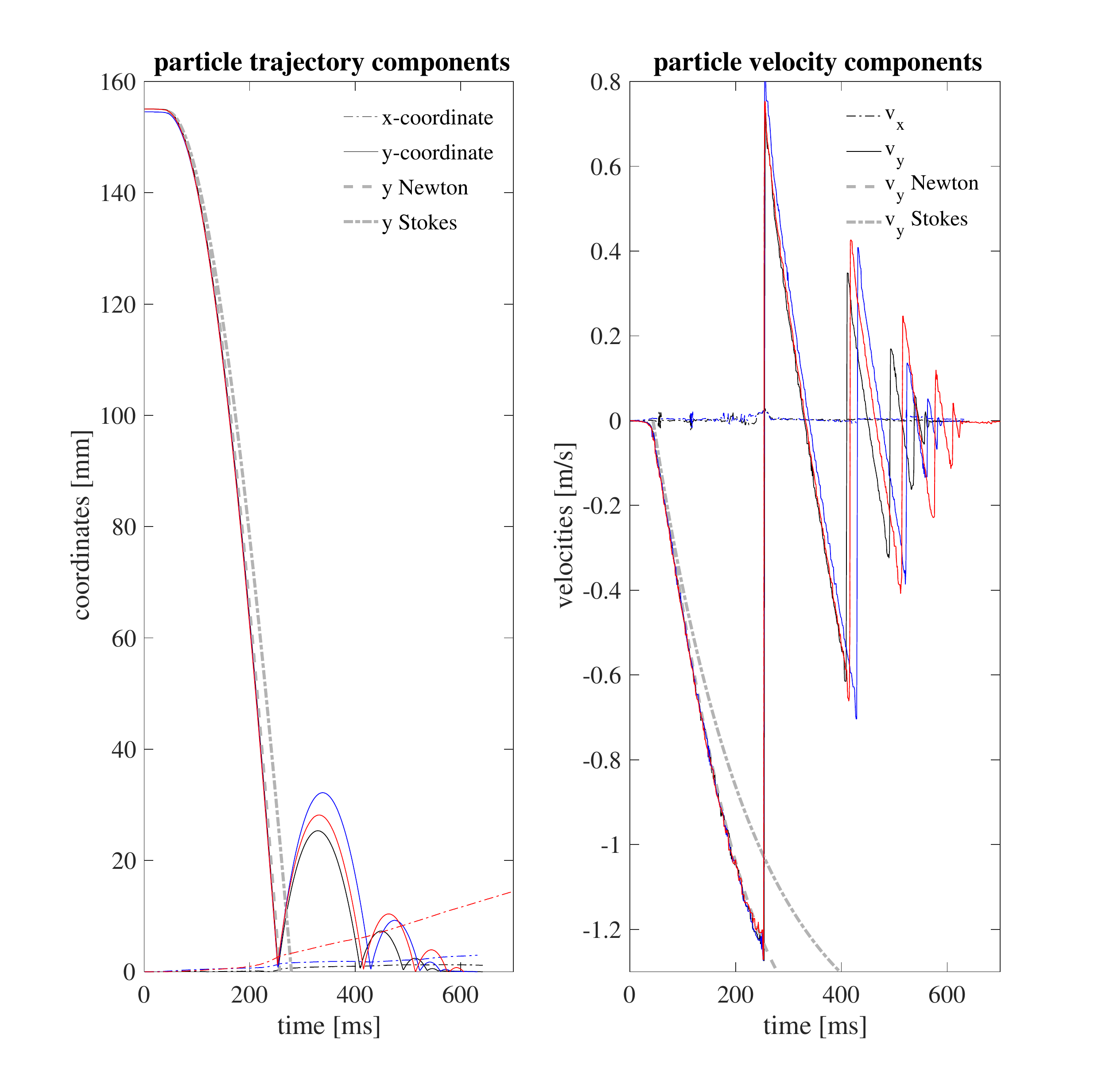}
\caption{Plots of particle trajectory (left) and velocity (right) over time for three independent settling experiments (shown with different colors -- blue, black, red) with the same steel sphere ($d_P=14.4~\unit{mm}$). Phase correction was applied to match the curves at the initial point. The lateral direction of motion can be obtained from the velocity plot, while the $x$-coordinate shows only the absolute value for better readability.}
\label{DiscFig2}
\end{figure}

\begin{figure}
\centering
\includegraphics[width=13cm]{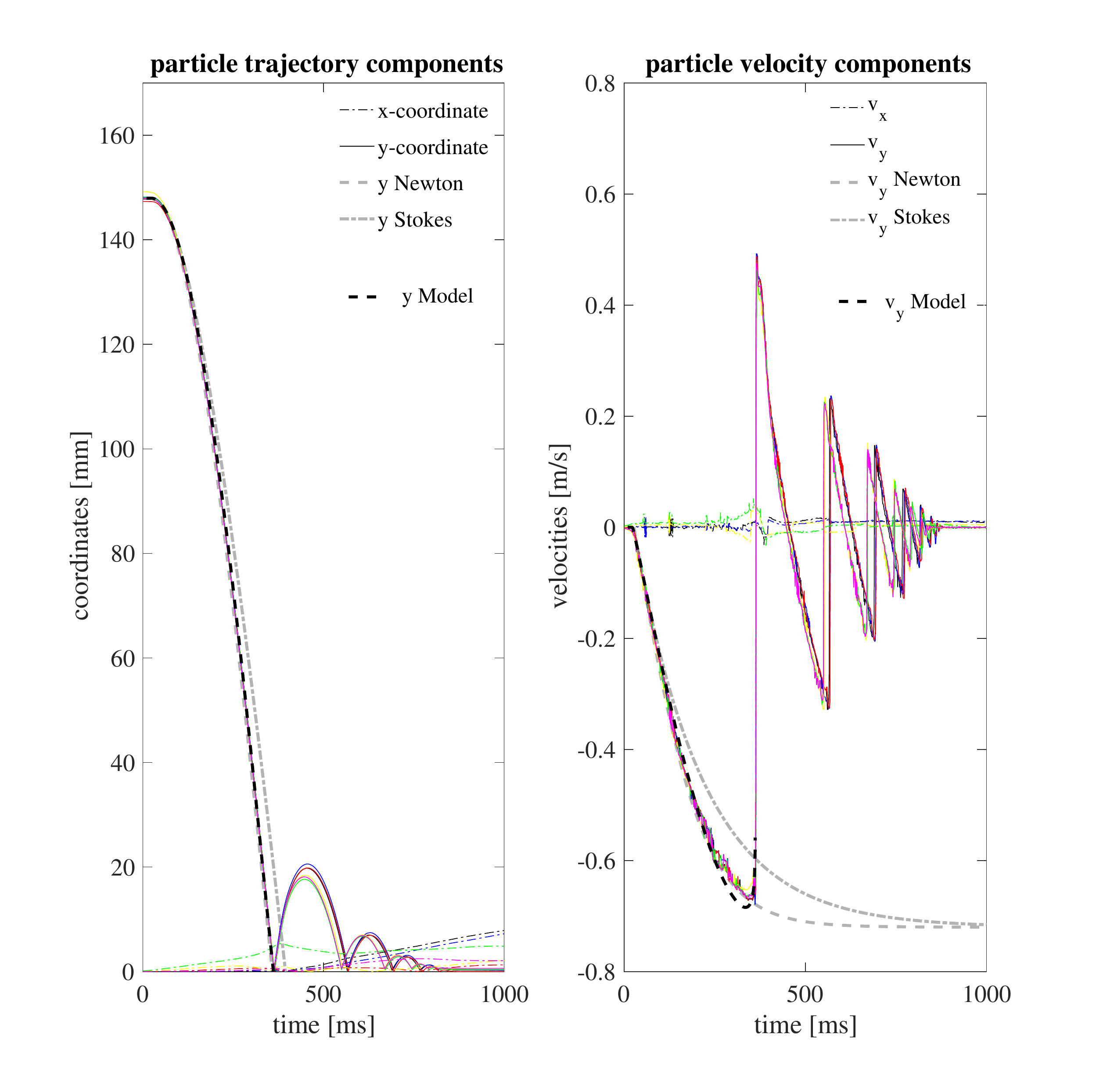}
\caption{Plots of particle trajectory (left) and velocity (right) over time for six independent settling experiments (shown with different colors) with the same PTFE sphere ($d_P=20~\unit{mm}$). Phase correction was applied to match the curves at the initial point. The lateral direction of motion can be obtained from the velocity plot, while the $x$-coordinate shows only the absolute value for better readability. Additionally, the numerical solution is shown as bold dashed line for both, the trajectory and the velocity. }
\label{DiscFig3}
\end{figure}

\begin{figure}
\centering
\includegraphics[width=13cm]{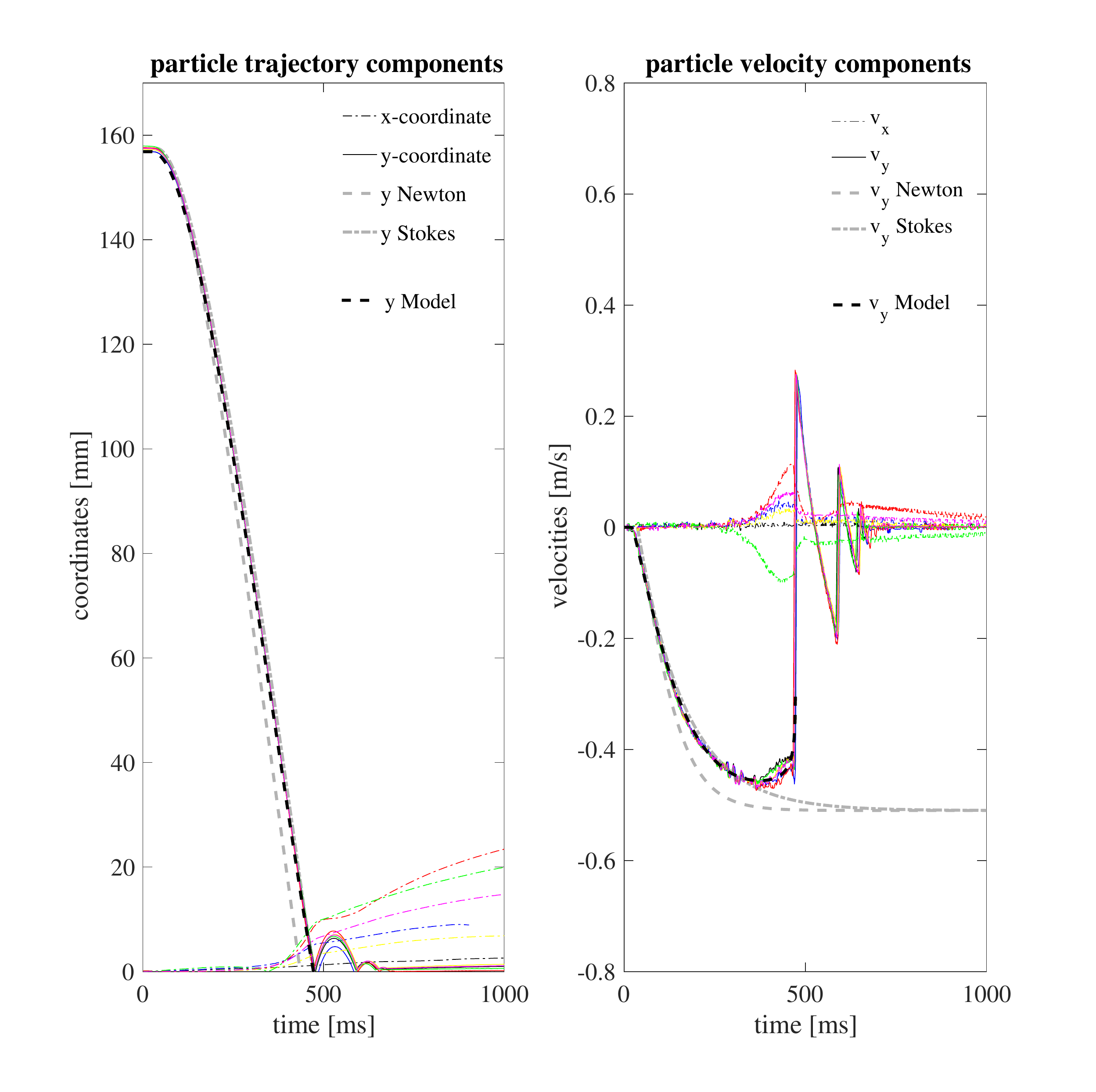}
\caption{Plots of particle trajectory (left) and velocity (right) over time for six independent settling experiments (shown with different colors) with the same PTFE sphere ($d_P=10~\unit{mm}$). Phase correction was applied to match the curves at the initial point. The lateral direction of motion can be obtained from the velocity plot, while the $x$-coordinate shows only the absolute value for better readability. Additionally, the numerical solution is shown as bold dashed line for both, the trajectory and the velocity.}
\label{DiscFig4}
\end{figure}

\begin{figure}
\centering
\includegraphics[width=13cm]{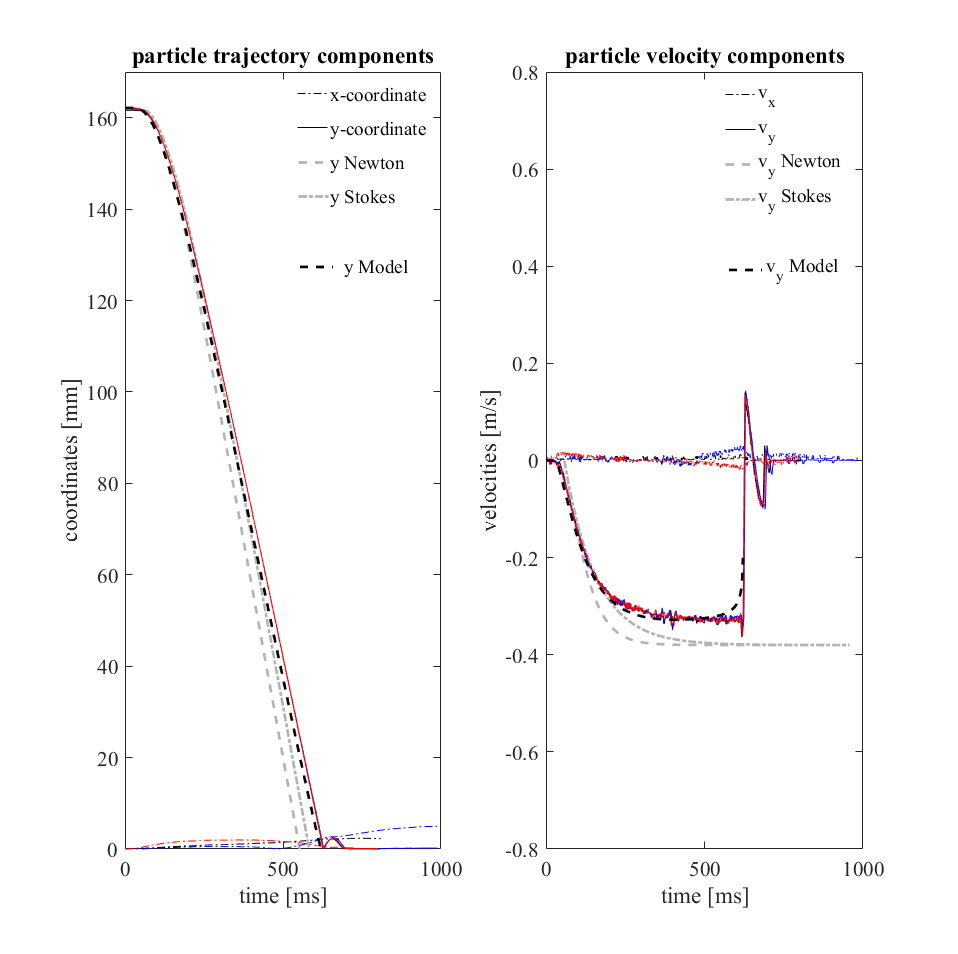}
\caption{Plots of particle trajectory (left) and velocity (right) over time for three independent settling experiments (shown with different colors -- blue, black, red) with the same PTFE sphere ($d_P=6~\unit{mm}$). Phase correction was applied to match the curves at the initial point. The lateral direction of motion can be obtained from the velocity plot, while the $x$-coordinate shows only the absolute value for better readability. Additionally, the numerical solution is shown as bold dashed line for both, the trajectory and the velocity.}
\label{DiscFig5}
\end{figure}

\begin{figure}
\centering
\includegraphics[width=13cm]{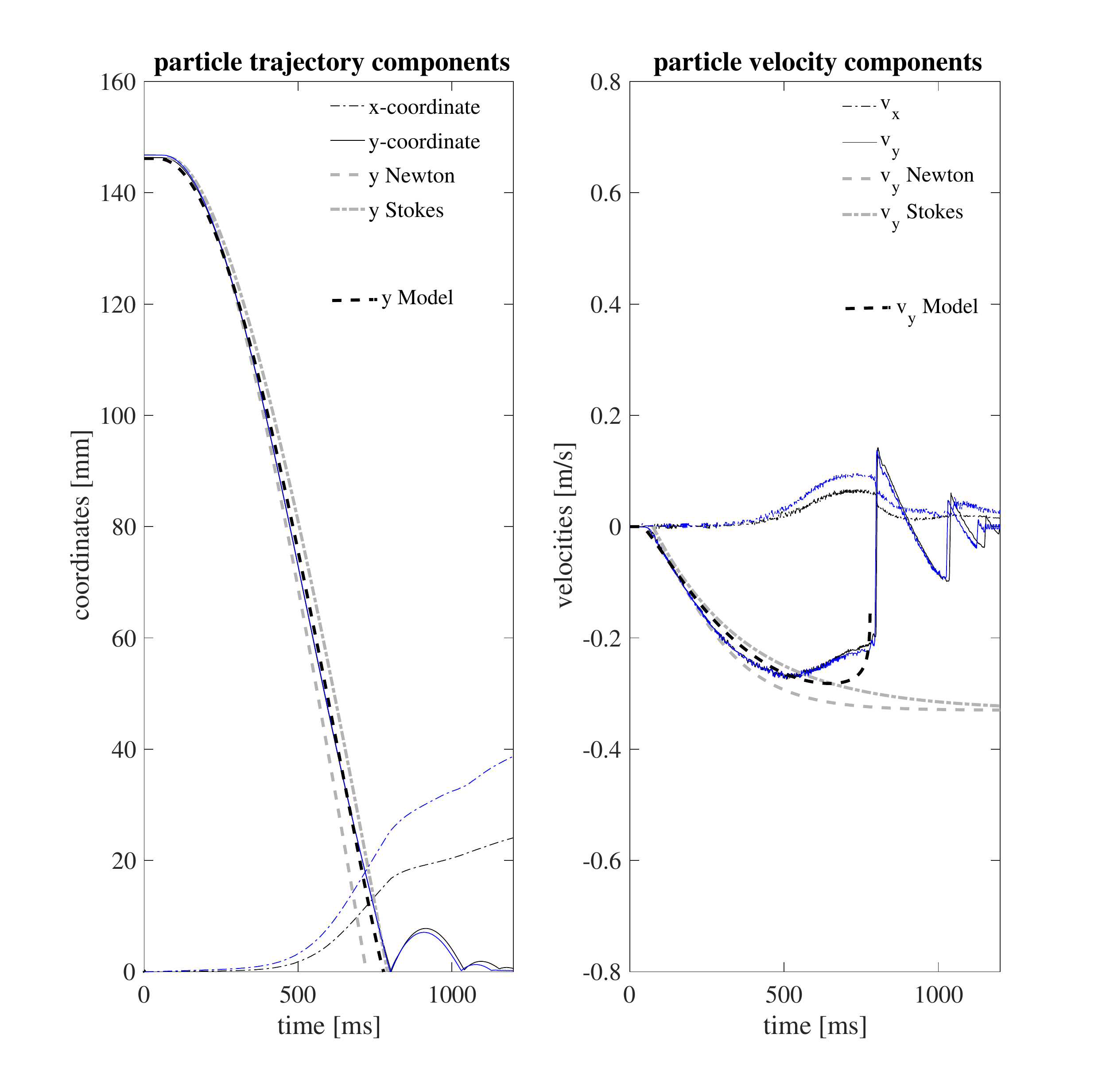}
\caption{Plots of particle trajectory (left) and velocity (right) over time for two independent settling experiments (shown with different colors -- blue and black) with the same POM sphere ($d_P=20~\unit{mm}$). Phase correction was applied to match the curves at the initial point. The lateral direction of motion can be obtained from the velocity plot, while the $x$-coordinate shows only the absolute value for better readability. Additionally, the numerical solution is shown as bold dashed line for both, the trajectory and the velocity.}
\label{DiscFig6}
\end{figure}

\begin{figure}
\centering
\includegraphics[width=13cm]{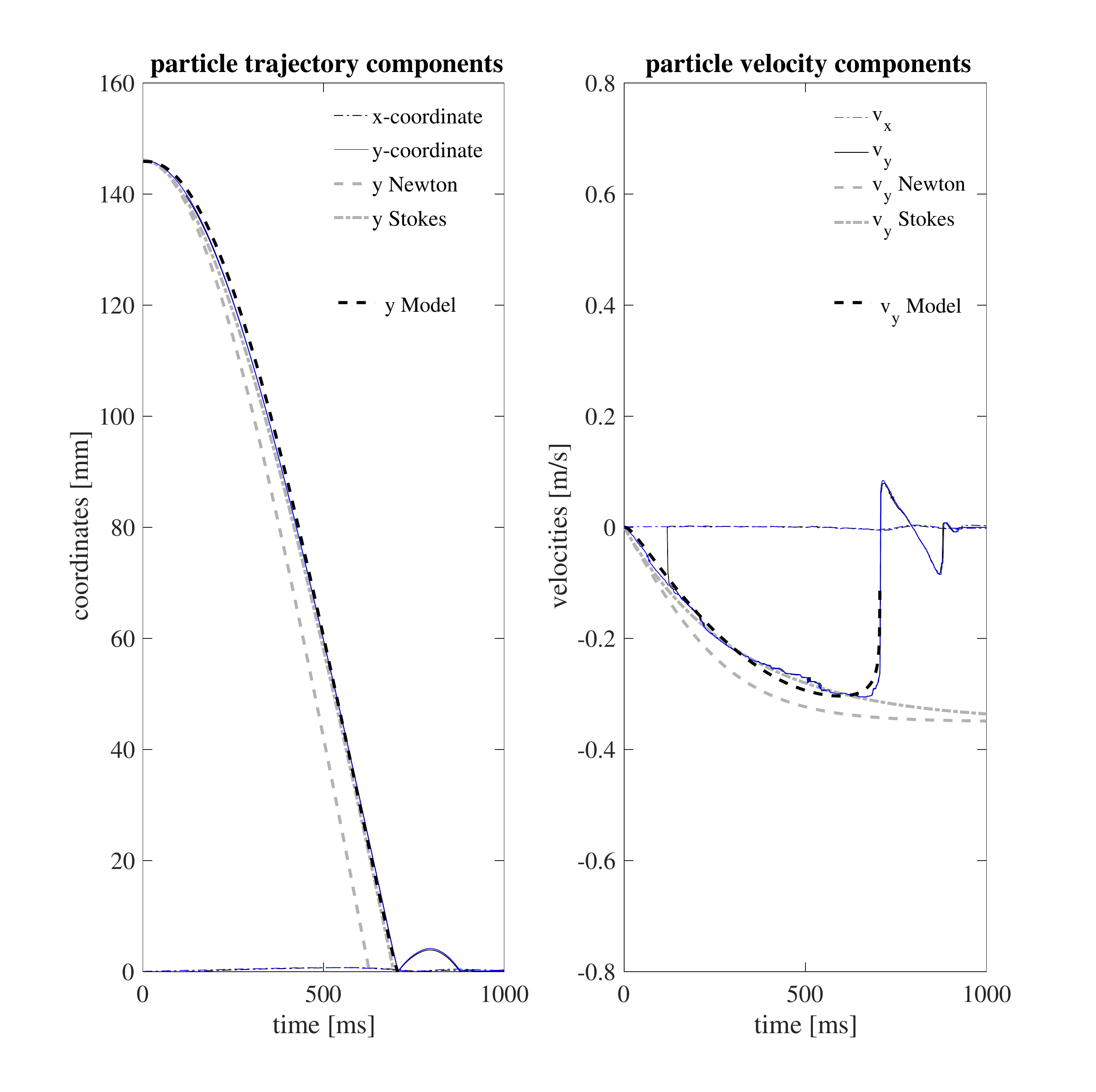}
\caption{Plots of particle trajectory (left) and velocity (right) over time for two independent settling experiments (shown with different colors -- blue and black) with the same rubber sphere ($d_P=22~\unit{mm}$). Phase correction was applied to match the curves at the initial point. The lateral direction of motion can be obtained from the velocity plot, while the $x$-coordinate shows only the absolute value for better readability. Additionally, the numerical solution is shown as bold dashed line for both, the trajectory and the velocity.}
\label{DiscFig7}
\end{figure}

 %

\end{document}